\documentclass[aps,prd,twocolumn,groupedaddress,showpacs,nofootinbib]{revtex4}
\usepackage{graphicx}

\begin{document}


\title{High-Energy Radiation from Remnants of Neutron Star Binary Mergers}

\author{Hajime Takami}
\email[]{takami@post.kek.jp}

\affiliation{JSPS Research Fellow -- Theory Center, Institute for Particle and Nuclear Studies, KEK, 1-1, Oho, Tsukuba 305-0801, Japan}

\author{Koutarou Kyutoku}
\email[]{kyutoku@uwm.edu}

\affiliation{JSPS Fellow for Research Abroad -- Department of Physics, University of Wisconsin-Milwaukee, P.O. Box 413, Milwaukee, Wisconsin 53201, USA}

\author{Kunihito Ioka}
\email[]{kunihito.ioka@kek.jp}

\affiliation{Theory Center, Institute for Particle and Nuclear Studies, KEK, 1-1, Oho, Tsukuba 305-0801, Japan, \\
Department of Particles and Nuclear Physics, the Graduate University for Advanced Studies (Sokendai), 1-1, Oho, Tsukuba 305-0801, Japan}

\date{\today}

\begin{abstract}
We study high-energy emission from the mergers of neutron star binaries as electromagnetic counterparts to gravitational waves aside from short gamma-ray bursts. The mergers entail significant mass ejection, which interacts with the surrounding medium to produce similar but brighter remnants than supernova remnants in a few years. We show that electrons accelerated in the remnants can produce synchrotron radiation in X-rays detectable at $\sim 100$ Mpc by current generation telescopes and inverse Compton emission in gamma rays detectable by the \emph{Fermi} Large Area Telescopes and the Cherenkov Telescope Array under favorable conditions. The remnants may have already appeared in high-energy surveys such as the Monitor of All-sky X-ray Image and the \emph{Fermi} Large Area Telescope as unidentified sources. We also suggest that the merger remnants could be the origin of ultra-high-energy cosmic rays beyond the knee energy, $\sim 10^{15}$ eV, in the cosmic-ray spectrum.
\end{abstract}

\maketitle 

\section{Introduction} \label{sec:introduction}

Recent development of gravitational wave (GW) interferometers such as advanced Laser Interferometer Gravitational wave Observatory (aLIGO) \cite{Abadie2010NIMPhysResSectA624p223}, advanced Virgo gravitational wave detector (AdV) \cite{Accadia2011CQGra28p114002}, and KAGRA [formerly called Large-scale Cryogenic Gravitational wave Telescope (LCGT)] \cite{Kuroda2010CQGra27p084004}, has raised expectations for detecting GWs from violent astrophysical events. The detection of GWs enables us to test general relativity and to open a new window to the universe.

The merger of compact binaries, including neutron stars (NSs) and/or black holes (BHs), is a primary source candidate of GWs. The detection rates of GWs from these binary mergers have been studied (e.g., \cite{Clark1979A&A72p120,Phinney1991ApJ380L17,Narayan1991ApJ379L17,Kochanek1993ApJ417L17}). Recent estimations suggest that the next-generation GW interferometers would detect $40$, $10$, and $20$ events per year for NS-NS, BH-NS, and BH-BH binaries, respectively (\cite{Abadie2010CQG27p173001} and references therein; see also \cite{Aasi2013arXiv1304.0670} about expectations for aLIGO and AdV in the forthcoming years). A more recent estimation suggests the detection rate of BH-BH binary mergers more than one order of magnitude higher \cite{Belczynski2012arXiv1208.0358}. The ongoing LIGO has already constrained the merger rates of these compact binaries. The upper limits on NS-NS, BH-NS, and BH-BH binary merger rates are $1.4 \times 10^{-2}$, $3.6 \times 10^{-3}$, and $7.3 \times 10^{-4}$ yr$^{-1} L_{10}^{-1}$ at 90\% confidence level, respectively \cite{Abbott2009PRD80p047101}. Here, $L_{10} = 10^{10} L_{{\rm B},\odot}$ and $L_{{\rm B},\odot} = 2.16 \times 10^{33}$ erg s$^{-1}$ is the solar luminosity in the B band \cite{Kopparapu2008ApJ675p1459}.

Recent simulations have revealed that the merger of neutron star binaries (NSBs), i.e., NS-NS and BH-NS binaries, leads to dynamical mass ejection. Newtonian and fully general relativistic simulations of NS-NS mergers have shown the subrelativistic ejection of NS material by tidal torque and/or shock heating with the mass of $M = 10^{-4} M_{\odot}$ - $10^{-2} M_{\odot}$ and the speed of $\beta = 0.1$ - $0.3$ in the unit of the speed of light $c$ for a wide range of parameters \cite{Rosswog1999A&A341p499,Ruffert2001A&A380p544,Hotokezaka2013PRD87p024001,Bauswein2013arXiv1302.6530}, where $M_{\odot}$ is the solar mass. Mass ejection via magnetically driven winds \cite{Shibata2011ApJ734L36} and neutrino-driven winds \cite{Dessart2009ApJ690p1681} may be also possible. The shock waves accelerated by a steep density gradient at a NS surface in the merging phase can also drive relativistic ejecta \cite{Kyutoku2012arXiv1209.5747}. A relativistic wind from a rapidly rotating and strongly magnetized NS produced in the post-merger stage may additionally inject energy to the ejecta, as discussed in the context of the extended emission of short gamma-ray bursts (SGRBs, e.g., \cite{Zhang2001ApJ552L35,Metzger2008MNRAS385p1455}). Mass ejection from BH-NS mergers has also been studied and the tidal mass ejection is also expected \cite{Lattimer1974ApJ192L145,Kyutoku2011PRD84p064018,Foucart2013PRD87p084006,Lovelace2013arXiv1302.6297,Deaton2013arXiv1304.3384,Kyutoku2013arXiv1305.6309}. The general relativistic simulations of BH-NS mergers show that the mass ejection is highly anisotropic with $M = 10^{-2} M_{\odot}$ - $10^{-1} M_{\odot}$ and $\beta = 0.2$ - $0.3$ \cite{Kyutoku2013arXiv1305.6309}.

Ejecta from NSBs are inevitably neutron-rich, and can drive r-process nucleosyntheses \cite{Lattimer1974ApJ192L145,Freiburghaus1999ApJ525L121}. The radioactive decay of r-process elements powers transient events, so-called macronovae or kilonovae \cite{Kulkarni2005astro-ph0510256,Li1998ApJ507L59,Metzger2010MNRAS406p2650,Barnes2013arXiv1303.5787}. Macronovae have been expected to be bright in optical bands at a few days after the mergers of NSBs \cite{Li1998ApJ507L59,Metzger2010MNRAS406p2650}. Recent studies with detailed opacity treatment including elements heavier than iron groups have indicated that macronovae are longer and softer events than those previously thought, over a week in infrared bands \cite{Barnes2013arXiv1303.5787,Tanaka2013arXiv1306.3742,Grossman2013arXiv1307.2943}. A recent near-infrared observation of short GRB 130603B by the Hubble Space Telescope (HST) at a post-burst phase finds radiation consistent with a macronova \cite{Berger2013arXiv1306.3960,Tanvir2013arXiv1306.4971}. Although this radiation can be also interpreted as an afterglow, this observation implies mass ejection with $M \sim 10^{-2} M_{\odot}$ and $\beta \sim 0.1$ if this is from a macronova.

The subsequent evolution of the subrelativistic ejecta is very similar to that of supernova remnants (SNRs). The ejecta expand freely in the initial phase \cite{Kyutoku2013arXiv1305.6309}. Then, they start to strongly interact with the interstellar medium (ISM) and to be decelerated when the mass of the swept-up ISM becomes comparable to the ejecta mass, entering into the Sedov-Taylor phase (e.g., \cite{Chevalier1982ApJ258p790}). The evolution of NSB remnants is generally faster than that of SNRs due to smaller ejected mass and the higher velocity of the ejecta. Note that the word of remnants is used for the remnants generated by interactions between the ejecta and the ISM, analogously to SNRs, not NSs or BHs produced by mergers throughout this paper. A fraction of the kinetic energy of the ejecta is converted to the energy of particles, i.e., the particles are accelerated at a forward shock in the remnants. In the magnetized medium, accelerated relativistic electrons radiate synchrotron photons in radio bands \cite{Meszaros1992ApJ397p570,Nakar2011Nature478p82,Piran2013MNRAS430p2121}.

In this paper, we show that relativistic electrons accelerated at the forward shocks of NSB remnants can also emit X-rays and gamma rays in the framework of a synchrotron self-Compton (SSC) model (e.g., \cite{Maraschi1992ApJ397L5,Bloom1996ApJ461p657}). Emission by SSC is inevitably present in a magnetized system with relativistic electrons. The ejecta of a NS-NS merger expand to the entire directions; NS material expands to the equatorial plane mainly due to tidal interaction, and it also expands along the rotational axis due to shock heating (e.g., \cite{Hotokezaka2013PRD87p024001}). The ratio of the velocities of the former and latter components is $\sim 2:1$ in fully general relativistic simulations \cite{Hotokezaka2013PRD87p024001}, while the former component is dominant, i.e., anisotropy is larger, in Newtonian simulations (i.e., \cite{Rosswog2000A&A360p171}) due to weaker gravity in the vicinity of the NSs than general relativity. In both cases, the expansion of the ejecta is reasonably approximated to be isotropic. A recent simulation revealing that radioactive heating quickly smooths out inhomogeneities in the initial mass distribution also supports this approximation \cite{Rosswog2013arXiv1307.2939}. Although NS-NS mergers could also produce the relativistic ejecta by the shock breakout \cite{Kyutoku2012arXiv1209.5747}, we consider only the subrelativistic ejecta in this study. On the other hand, the ejecta of a BH-NS merger expand anisotropically at the initial phase. Then, when they are decelerated by interactions with the ISM, they start to expand in a nearly isotropic manner because their lateral expansion speed is roughly the sound speed comparable with the speed of the shock. Thus, we emphasize that the model based on the approximation of the isotropic expansion in this paper is also applicable to the cases of BH-NS mergers.

It is worth noting that there are few models for X-ray and higher energy emission associated with GWs. Isotropic emission is eagerly anticipated to find electromagnetic counterparts that help the localization of GW sources because the SGRB jet, if any, is off-axis in most cases and not able to be detected. X-ray emission in the inspiral phase is proposed, which may be precursors of SGRBs \cite{Ioka2000ApJ537p327,Hansen2001MNRAS322p695,Tsang2012PRL108p011102}. The relativistic ejecta driven by accelerating shock waves generated just after the collision of NSs produce non-thermal electrons via shock acceleration, emitting synchrotron radiation \cite{Kyutoku2012arXiv1209.5747}. The synchrotron emission reaches X-ray energies and is radiated almost isotropically because the ejecta expand nearly isotropically. Another model is the early "afterglow" of a NS-NS merger lasting several thousand seconds powered by the dissipation of the kinetic energy of the magnetized wind from a rapidly-spinning massive NS \cite{Zhang2013ApJ763L22}. This model predicts a large opening angle of emission compared to SGRBs ($30^{\circ}$--$40^{\circ}$ for a reasonable parameter choice \cite{Bucciantini2012MNRAS419p1537}), which is determined by the balance between the pressure of the magnetar wind nebula and ejecta. Also, high-energy neutrinos from magnetar wind nebulae are proposed \cite{Gao2013arXiv1306.3006}. Compared to these models our model is based on more conservative mass ejection, i.e., dynamical mass ejection, confirmed by numerical simulations. Also, this study is the first estimation of inverse-Compton-scattered photons for isotropic radiation.

This paper is laid out as follows. We start with the description of our model in Section \ref{sec:model}. Then, we show the spectral energy distribution (SED) of NSB merger remnants (NSBMRs) and investigate their dependence on parameters related to particle acceleration in Section \ref{sec:results}. The detectability of the emission is discussed with particular attention to possible backgrounds in Section \ref{sec:detectability}. Then, we discuss the dependence of the SED on parameters of the environments and the ejecta properties of the merger, and examine possible hadronic emission in Section \ref{sec:discussion}. Then, this study is summarized in Section \ref{sec:summary}.

\section{The Model} \label{sec:model}

Consider mass $M$ dynamically ejected isotropically from a NS-NS merger with the speed of $\beta c$. The ejecta initially expand freely and isotropically, and sweep up the surrounding ISM with the number density $n$. When the total mass of the swept ISM becomes comparable with $M$, the ejecta are started to be decelerated and the expansion enters into the Sedov-Taylor phase. A strong forward shock is generated where particles are accelerated. The deceleration radius is 
\begin{equation}
R_{\rm dec} = \left( \frac{3 M}{4 \pi n m_p} \right)^{1/3} = 1 \times 10^{18} M_{-2}^{1/3} n_0^{-1/3} ~~~{\rm cm},
\label{eq:rdec}
\end{equation}
and the corresponding deceleration time $t_{\rm dec}$ is 
\begin{equation}
t_{\rm dec} = \frac{R_{\rm dec}}{\beta c} = 5 M_{-2}^{1/3} n_0^{-1/3} \beta_{0.3}^{-1} ~~~{\rm yr},
\label{eq:tdec}
\end{equation}
where $M_{-2} = M / (10^{-2} M_{\odot})$, $n_0 = n / 10^0$ cm, $\beta_{0.3} = \beta / 0.3$, and $m_p$ is the proton mass. In the case of a BH-NS merger, the deceleration radius and time may be modified by a factor of $\left( 4 \pi / \Delta \Omega \right)^{1/3}$ where $\Delta \Omega$ is the solid angle where NS material is initially ejected. This factor is small, only $\sim 2$ according to general relativistic simulations \cite{Kyutoku2013arXiv1305.6309}. Thus, although the calculation results below assume isotropic mass ejection, they can be reasonably applied to the BH-NS merger cases just by replacing parameters with those of BH-NS mergers. Throughout this paper we estimate emission from accelerated electrons at $t = t_{\rm dec}$ when it is the most luminous.

Let us roughly estimate the bolometric flux of emission from a NSBMR occurred at $D = 100$ Mpc from the Earth. We assume that relativistic electrons have a fraction $\epsilon_e$ of the total kinetic energy of the ejecta $E = M \beta^2 c^2 / 2$. If the energy of the electrons is converted into radiation at $t = t_{\rm dec}$, the bolometric flux is 
\begin{eqnarray}
f_{\rm max} &=& \frac{\epsilon_e E}{4 \pi D^2 t_{\rm dec}} \nonumber \\
&\simeq& 4 \times 10^{-13} 
\epsilon_{e,-1} M_{-2}^{2/3} \beta_{0.3}^3 n_{0}^{1/3} \nonumber \\
&& ~~~~~~~~~~~~ \times D_{2}^{-2} ~~{\rm erg~cm}^{-2}~{\rm s}^{-1},
\label{eq:fmax}
\end{eqnarray}
where $\epsilon_{e,-1} = \epsilon_e / 10^{-1}$ and $D_2 = D / 10^2$ Mpc. This numerical value is well above the sensitivity of active X-ray telescopes such as \emph{Chandra} X-ray observatory, the \emph{X-ray Multi-mirror Mission Newton} (\emph{XMM}-\emph{Newton}), \emph{Suzaku} X-ray telescope in soft X-ray ranges for point-like sources. Also, it is comparable with the sensitivity of the Cherenkov Telescope Array (CTA) in the very-high-energy ($> 100$ GeV) range \cite{Actis2011ExA32p193}. The distance of $100$ Mpc is within the horizon of aLIGO and AdV to NS-NS and BH-NS binaries \cite{Abadie2010CQG27p173001}. Given a NS-NS merger rate per galaxy of $\sim 10^{-4}$ yr$^{-1}$ from known NS-NS mergers in the Milky Way and the number density of galaxies of $\sim 10^{-2}$ Mpc$^{-3}$, the number of NS-NS binary merger remnants within the distance of $D$ can be estimated as 
\begin{eqnarray}
N_{\rm MR} &\sim& \left( 10^{-4} {\rm yr}^{-1} \right) 
\left( 10^{-2} {\rm Mpc}^{-3} \right) \frac{4\pi}{3} D^3 t_{\rm dec} \nonumber \\
&\sim& 21 D_2^3 M_{-2}^{1/3} n_0^{-1/3} \beta_{0.3}^{-1}. 
\label{eq:nmr}
\end{eqnarray}
Therefore, high-energy emission from NSBMRs may become good observational targets for these instruments if they produce photons in the high-energy bands. Thus, we consider NSBMRs within 100 Mpc throughout this paper that allow us to neglect cosmological effects. The detection of such high-energy photons can confirm that electrons are accelerated.

We calculate the broadband SED of NSBMRs in the framework of a SSC model with a numerical code developed in \cite{Takami2011MNRAS413p1845}. In the SSC model synchrotron radiation and SSC emission are calculated based on a given electron spectrum and the strength of magnetic fields. The inverse Compton scattering (ICS) of cosmic microwave background photons and bremsstrahlung are negligible in the range surveyed in this study. The model parameters defined in this Section are summarized in Table \ref{tab:parameters}. The table also exhibits the fiducial parameters used in this study for NS-NS merger remnants. In the cases of BH-NS merger remnants, a larger value of the ejected mass $M$ ($\sim 0.1 M_{\odot}$) can be allowed \cite{Kyutoku2013arXiv1305.6309}.

We assume that a fraction $\epsilon_B$ of internal energy released at the forward shock is converted to the energy of magnetic fields downstream of the shock. Rankine-Hugoniot relations indicate that the released internal energy density is $9 n m_p \beta^2 c^2 / 8$ for a non-relativistic gas, and therefore the strength of the magnetic fields is 
\begin{equation}
B = \left( 9 \pi \epsilon_B n m_p \right)^{1/2} \beta c = 6 \times 10^{-3} \epsilon_{B,-2}^{1/2} n_0^{1/2} \beta_{0.3} ~~~{\rm G}, 
\label{eq:mag}
\end{equation}
where $\epsilon_{B,-2} = \epsilon_B / 10^{-2}$.

The spectrum of accelerated electrons at injection is assumed to be a power-law function with the index of $s$ in the range from the minimum Lorentz factor at injection $\gamma_m$ and the maximum Lorentz factor of electrons $\gamma_{\rm max}$ following diffusive shock acceleration \cite{Drury1983RPPh46p973,Blandford1987PhR154p1}, 
\begin{equation}
\frac{d N}{d\gamma}(\gamma) = \tilde{N} \gamma^{-s} ~~~( \gamma_m < \gamma < \gamma_{\rm max} ), 
\end{equation}
where $\gamma$ is the Lorentz factor of electrons and $\tilde{N}$ is the normalization factor of the total number of electrons.

The maximum Lorentz factor of electrons can be estimated by equating the time scale of particle acceleration and the minimum time scale among the dynamical timescale $t_{\rm dyn} = R_{\rm dec} / \beta c~(= t_{\rm dec}$ in this study$)$ and cooling time scales. The acceleration time scale is estimated in the framework of diffusive shock acceleration for a parallel shock in the test particle approximation as (e.g., \cite{Drury1983RPPh46p973,Blandford1987PhR154p1}) 
\begin{equation}
t_{\rm acc} = \frac{20 \xi r_{\rm L}}{3 \beta^2 c}, 
\end{equation}
and the Larmor radius of the particles $r_{\rm L}$ is 
\begin{equation}
r_{\rm L} = \frac{\gamma m_e c^2}{e B}, 
\end{equation}
where $m_e$ is the electron mass, $e$ is the absolute electric charge of an electron, and $\xi \geq 1$ is called a gyrofactor, representing particle acceleration efficiency; a larger $\xi$ results in less efficient acceleration because particles require more time to go back and forth between the downstream and upstream of the shock. Although X-ray observations have revealed $\xi \sim 1$ in the cases of shell-type SNRs \cite{Yamazaki2004A&A416p595,Parizot2006A&A453p387}, it is still difficult to predict the value of $\xi$ from theories. Thus, we treat $\xi$ as a free parameter.

Electrons are cooled predominantly by synchrotron radiation with the time scale of 
\begin{equation}
t_{\rm syn} = \frac{6 \pi m_e c}{\sigma_T \gamma B^2}, 
\label{eq:tsyn}
\end{equation}
where $\sigma_T$ is the cross-section of Thomson scattering. Therefore, the maximum Lorentz factor of electrons is 
\begin{equation}
\gamma_{\rm max} = {\rm min} \left( 
\frac{3 e \beta R_{\rm dec} B}{20 \xi m_e c^2},~
\left[ \frac{9 \pi e \beta^2}{10 \xi \sigma_T B} \right]^{1/2} 
\right). 
\label{eq:gmax}
\end{equation}
When this is regarded as a function of $\xi$, this expression is numerically 
represented as 
\begin{eqnarray}
\gamma_{\rm max} = \left\{ 
\begin{array}{ll}
2 \times 10^{8} \xi^{-1/2} \epsilon_{B,-2}^{-1/4} n_0^{-1/4} \beta_{0.3}^{1/2} & 
( \xi < \xi_{\rm b} ) \\
2 \times 10^{11} \xi^{-1} \epsilon_{B,-2}^{1/2} M_{-2}^{1/3} n_0^{1/6} 
\beta_{0.3}^2 & ( \xi \geq \xi_{\rm b} ), 
\end{array}
\right.
\label{eq:gmaxnum}
\end{eqnarray}
where 
\begin{equation}
\xi_{\rm b} = \frac{\sigma_T e B^3 R_{\rm dec}^2}{40 \pi m_e^2 c^4} = 
2 \times 10^6 \epsilon_{B,-2}^{3/2} M_{-2}^{2/3} n_0^{5/6} \beta_{0.3}^3.
\end{equation}
The particle acceleration is limited by synchrotron cooling for $\xi < \xi_{\rm b}$ and it is restricted by the dynamical time $t_{\rm dec}$ for $\xi \geq \xi_{\rm b}$. In many cases the maximum Lorentz factor $\gamma_{\rm max}$ is limited by the synchrotron cooling in this study.

The normalization factor $\tilde{N}$ and the minimum Lorentz factor of electrons at injection $\gamma_m$ are determined from the total energy and number of accelerated electrons. The definition of $\epsilon_e$ is expressed as 
\begin{equation}
\epsilon_e E = \tilde{N} m_e c^2 \int_{\gamma_m}^{\gamma_{\rm max}} d\gamma \gamma^{1-s}. 
\label{eq:energy}
\end{equation}
On the other hand, assuming that only a fraction $\eta$ of bulk electrons are accelerated, the number conservation is represented as 
\begin{equation}
\eta N = \tilde{N} \int_{\gamma_m}^{\gamma_{\rm max}} d\gamma \gamma^{-s}. 
\label{eq:number}
\end{equation} 
Here, the total number of bulk electrons $N$ is $4 \pi R_{\rm dec}^3 n / 3$. In order to estimate $\gamma_m$ and $\tilde{N}$,  Eqs. (\ref{eq:energy}) and (\ref{eq:number}) are solved numerically by the Newton-Raphson method. Note that $\gamma_m$ is written down approximately in an analytical form if $s > 2$ and $\gamma_{\rm max}$ is high enough, 
\begin{equation}
\gamma_m \simeq \frac{\epsilon_e E}{\eta N m_e c^2} \frac{s - 2}{s - 1}. 
\label{eq:appgmin}
\end{equation}
This expression is useful to understand the dependence of $\gamma_m$ on parameters, such as on the fraction of accelerated electrons $\eta$.

Assuming the continuous injection of electrons with a flat rate, synchrotron cooling increases the spectral index by unity above a characteristic Lorentz factor $\gamma_c$. This cooling Lorentz factor $\gamma_c$ is estimated by equating $t_{\rm syn}$ and $t_{\rm dyn}$, 
\begin{equation}
\gamma_c = \frac{6 \pi \beta m_e c^2}{\sigma_T B^2 R_{\rm dec}} = 1 \times 10^5 \epsilon_{B,-2}^{-1} M_{-2}^{-1/3} n_{0}^{-2/3} \beta_{0.3}^{-1}. 
\label{eq:gc}
\end{equation}
Therefore, in a slow cooling regime (i.e., $\gamma_m < \gamma_c$) the electron spectrum at $t = t_{\rm dec}$ is 
\begin{eqnarray}
\frac{d N}{d\gamma}(\gamma) = \left\{ 
\begin{array}{ll}
\tilde{N} \gamma^{-s} & ( \gamma_m < \gamma < \gamma_c ) \\
\tilde{N} \gamma_c \gamma^{-s-1} & ( \gamma_c < \gamma < \gamma_{\rm max} )
\end{array}
\right., 
\end{eqnarray}
while in a fast cooling regime (i.e., $\gamma_m > \gamma_c$), 
\begin{eqnarray}
\frac{d N}{d\gamma}(\gamma) = \left\{ 
\begin{array}{ll}
\tilde{N} \gamma_c \gamma_m^{1-s} \gamma^{-2} & ( \gamma_c < \gamma < \gamma_m ) \\
\tilde{N} \gamma_c \gamma^{-s-1} & ( \gamma_m < \gamma < \gamma_{\rm max} )
\end{array}
\right..
\end{eqnarray}

There are a few cases that electrons are cooled predominantly by ICS rather than synchrotron radiation in the parameter range investigated in this study. In these cases $\gamma_c$ is replaced by the cooling Lorentz factor estimated by the ICS energy-loss time scale, $t_{\rm ICS}$. In the Thomson regime the ICS cooling time scale is evaluated by replacing the energy density of magnetic fields in Eq. (\ref{eq:tsyn}) by the energy density of photons. Since the Klein-Nishina (KN) effect becomes significant in some cases, we adopt the formula of ICS cooling applicable in both Thomson and KN regimes in an isotropic photon field to estimate $\gamma_{\rm max}$ and $\gamma_c$, 
\begin{eqnarray}
t_{\rm ICS} &=& \frac{4}{3 \sigma_T c} \left[ 
\int d\epsilon_{\gamma} \frac{1}{b \epsilon_{\gamma}} 
\frac{d n}{d\epsilon_{\gamma}} 
\left\{ \left( 6 + \frac{b}{2} + \frac{6}{b} \right) \log \left( 1 + b \right) 
\right. \right. \nonumber \\
&& - \log^2 \left( 1 + b \right) 
- 2 {\rm Li}_2 \left( \frac{b}{1+b} \right) \nonumber \\
&& \left. \left. 
- \frac{1}{(1 + b)^2} 
\left( \frac{11}{12} b^3 + 8 b^2 + 13 b + 6 \right) 
\right\}
\right]^{-1}, 
\end{eqnarray}
where $b = 4 \gamma \epsilon_{\gamma}$ and $\epsilon_{\gamma}$ is the energy of photons in the unit of $m_e c^2$ \cite{Aharonian1981Ap&SS79p321}. The dilogarithm is defined as 
\begin{equation}
{\rm Li}_2(x) \equiv \int_x^0 dt \frac{\log (1 - t)}{t}. 
\end{equation}
When the ICS cooling of electrons is dominant, the electron spectrum deviates from a simple broken power-law function in a high-energy range where the KN effect is significant. A spectrum becomes generally harder in the KN regime than in the Thomson regime because of the weaker energy-loss rate of electrons. Thus, the approximation of a broken power-law function provides conservative flux estimations in the energy range where the KN effect is significant.

In practice, we calculate the SED of NSBMRs iteratively. First, it is calculated on the assumption that the synchrotron energy-loss of electrons is dominant. The resultant SED is utilized to evaluate the ICS energy-loss of electrons. Then, the SED is re-calculated with the ICS loss. The SED of NSBMRs is iteratively calculated until the radiation energy density converges within 5\%.

\begin{table}
\caption{Model parameters}
\label{tab:parameters}
\begin{ruledtabular}
\begin{tabular}{lcc}
& Symbols & Fiducial \\
\hline
Ejected mass & $M$ & $10^{-2} M_{\odot}$ \\
Initial speed of ejecta & $\beta$ & $0.3$ \\
Number density of ambient matter & $n$ & $1$ cm$^{-3}$ \\
Energy fraction of electrons & $\epsilon_e$ & $0.1$ \\
Energy fraction of magnetic fields  & $\epsilon_B$ & $10^{-2}$ \\
Spectral index & $s$ & $2.2$ \\
Gyrofactor & $\xi$ & $1$ \\
Fraction of accelerated electrons & $\eta$ & $1$ \\
Distance & $D$ & $100$ Mpc
\end{tabular}
\end{ruledtabular}
\end{table}

\section{Spectral Energy Distribution} \label{sec:results}

\subsection{Fiducial Case}

\begin{figure}
\includegraphics[clip,width=0.95\linewidth]{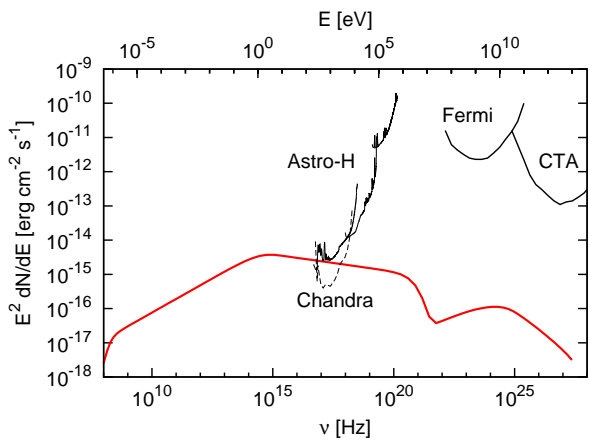}
\caption{SED of a NSBMR at $t = t_{\rm dec}$ and $D = 100$ Mpc with the fiducial model parameters in Table \ref{tab:parameters}. The differential sensitivity curves of X-ray and gamma-ray instruments are also indicated: \emph{Chandra} and Astro-H (100 ks; http://astro-h.isas.jaxa.jp/researchers/sim/sensitivity.html), the \emph{Fermi}-LAT (1 year for the survey mode) \cite{Atwood2009ApJ697p1071}, and CTA (50 hours; the goal sensitivity) \cite{Actis2011ExA32p193}.}
\label{fig:fid}
\end{figure}

Figure \ref{fig:fid} shows the SED of a NSBMR at $t = t_{\rm dec}$ and $D = 100$ Mpc calculated with our fiducial parameters in Table \ref{tab:parameters}. In this case electrons are in the slow cooling regime and the synchrotron cooling of electrons is dominant. The differential sensitivity curves of \emph{Chandra}, ASTRO-H, the \emph{Fermi} Large Area Telescope (LAT), and CTA are also plotted. Note that the differential sensitivity of \emph{XMM}-\emph{Newton} is roughly $50\%$ better than that of \emph{Chandra} though it is not plotted.

The lower energy component ($\lesssim 10^{21}$ Hz) consists of synchrotron radiation. The SED of the synchrotron radiation peaks at $\nu = \nu_{\rm p,syn}$ in an optical band as analytically estimated as \cite{Crusius1986AA164L16}
\begin{eqnarray}
\nu_{\rm p,syn} &=& 0.58~\frac{3 \gamma_c^2 e B}{2 \pi m_e c} \nonumber \\
&=& 5 \times 10^{14} \epsilon_{B,-2}^{-3/2} M_{-2}^{-2/3} n_0^{-5/6} \beta_{0.3}^{-1} ~~{\rm Hz}, 
\label{eq:syn_brk}
\end{eqnarray}
as long as electrons are predominantly cooled by synchrotron radiation (Eq. \ref{eq:gc}) in the slow cooling regime and $2 < s < 3$. The coefficient of $0.58$ is a slight modification to the characteristic frequency of synchrotron radiation, which is obtained by a numerical calculation that the SED of the synchrotron radiation of an electron with certain energy peaks at $0.58$ times the characteristic frequency. The roll-off frequency of the synchrotron radiation $\nu_{\rm syn}^{\rm max}$ is also analytically estimated in the same way by replacing $\gamma_c$ by $\gamma_{\rm max}$, 
\begin{eqnarray}
\nu_{\rm syn}^{\rm max} = \left\{ 
\begin{array}{ll}
9 \times 10^{20} \xi^{-1} \beta_{0.3}^2 ~~{\rm Hz} & ( \xi < \xi_{\rm b} ) \\
2 \times 10^{27} \xi^{-2} \epsilon_{B,-2}^{3/2} M_{-2}^{2/3} n_0^{5/6} 
\beta_{0.3}^5 ~~{\rm Hz} & ( \xi \geq \xi_{\rm b} ), 
\end{array}
\right.
\label{eq:syn_max}
\end{eqnarray}
when $\nu_{\rm syn}^{\rm max}$ is regarded as a function of $\xi$. Since the roll-off frequency is limited by synchrotron cooling here, the upper formula is applied. The numerical calculation reproduces these analytical estimations very well. The effect of synchrotron self-absorption appears below $\sim 2 \times 10^8$ Hz.

The higher energy component ($\gtrsim 10^{21}$ Hz) results from the SSC emission of electrons. The photon energy at the SED peak of the SSC emission is 
\begin{eqnarray}
E_{\rm p,SSC} &\sim& \frac{4}{3} \gamma_c^2 h \nu_{\rm p,syn} \nonumber \\
&=& 5 \times 10^{10} \epsilon_{B,-2}^{-7/2} M_{-2}^{-4/3} n_0^{-13/6} \beta_{0.3}^{-3} ~~{\rm eV}, 
\label{eq:icspeak}
\end{eqnarray}
where $h$ is Planck constant. This is described in the unit of eV according to convention of gamma-ray communities. This value is also reproduced in the numerical calculation.

In figure \ref{fig:fid} the flux of synchrotron radiation is higher than the sensitivity curves of \emph{Chandra} and Astro-H in soft X-ray bands, while the SSC emission is far from the sensitivity of the gamma-ray observatories. Note that the sensitivity of Astro-H is the highest at its lowest energy range. Therefore, NSBMRs are detectable in X-rays bands under the fiducial parameter set. However, the flux of the SED peak is $\sim 4 \times 10^{-15}$ erg cm$^{-2}$ s$^{-1}$, being two orders of magnitude smaller than the maximally achievable bolometric flux $f_{\rm max}$ (Eq. \ref{eq:fmax}). This is because 1) only electrons with $\gamma > \gamma_c$ efficiently radiate while the lower energy electrons have larger energy due to $s > 2$ and 2) the SED spreads over a wide range. The ratio of the energies in electrons above and below $\gamma_c$ can be roughly estimated as $[\gamma_c^2 dN / d\gamma (\gamma_c)] / [\gamma_m^2 dN / d\gamma (\gamma_m)] \simeq 0.1$. Also, the radiation is distributed over more than ten orders of magnitude in frequency, which reduces the flux in a band by $\sim 0.1$. These two effects explain the two-order-of-magnitude difference. 

\subsection{Parameter dependence on particle acceleration}

\begin{figure}
\includegraphics[clip,width=0.95\linewidth]{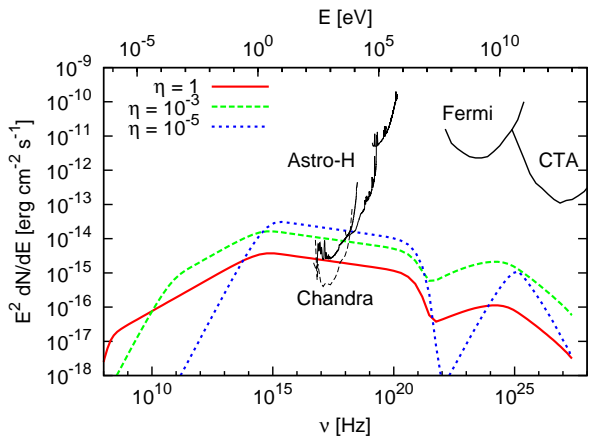}
\caption{Dependence of the NSBMR SED at $t = t_{\rm dec}$ on the number fraction of accelerated electrons $\eta$. The adopted values of $\eta$ are indicated in the legend. The other parameters are the fiducial values. The red solid line corresponds to the fiducial case shown in figure \ref{fig:fid}.}
\label{fig:eta}
\end{figure}

The fraction of electrons participating in particle acceleration $\eta$ can be less than unity in analogy with SNRs. It has commonly believed that a part of bulk electrons are accelerated in SNRs. However, the fraction of accelerated electrons is difficult to firmly predict from particle acceleration theories at present. Thus, $\eta$ is treated as a free parameter and its dependence is examined.

Figure \ref{fig:eta} shows the dependence of the NSBMR SED at $t = t_{\rm dec}$ and $D = 100$ Mpc on the number fraction of accelerated electrons $\eta$ to the bulk electrons. In all the cases electrons are primarily cooled by synchrotron radiation. As indicated in Eq. (\ref{eq:appgmin}), a smaller value of $\eta$ leads to a higher minimum Lorentz factor of electrons at injection $\gamma_m$. The absence of low-energy electrons due to high $\gamma_m$ results in 1) the reduction of flux in radio bands and 2) the enhancement of radiative efficiency due to the relative increase of electrons with $\gamma \gtrsim \gamma_c$. The value of $\gamma_m$ and underlying parameter $\eta$ affects the detectability of radio emission discussed in \cite{Piran2013MNRAS430p2121}.

In the intermediate $\eta$ case ($\eta = 10^{-3}$), the spectral break at $\sim 10^{11}$ Hz is no longer due to synchrotron self-absorption but due to the minimum Lorentz factor of electrons $\gamma_m$. This can be confirmed by the spectral slope proportional to $\nu^{4/3}$ below the break frequency. Note that the SED of synchrotron radiation below the frequency of synchrotron self-absorption is proportional to $\nu^{3/2}$. We can also find the increase of radiation flux by the enhancement of radiative efficiency in both synchrotron radiation and SSC emission. Recall that the target photons of SSC emission are synchrotron photons; the number of SSC photons with certain energy is roughly proportional to the squared number of the electrons emitting the SSC photons. Electrons are still in the slow cooling regime.

When the accelerated electron fraction $\eta$ is very small and reaches $\eta = 10^{-5}$, the minimum Lorentz factor at injection $\gamma_m$ exceeds the Lorentz factor of the cooling break $\gamma_c$ and electrons enter the fast cooling regime. In this regime, the frequency of the spectral break originating from $\gamma_m$ is higher than that produced by $\gamma_c$, opposite to the cases in the slow cooling regime, although it is difficult to distinguish these two breaks by eye in figure \ref{fig:eta}, as they are very close at $\sim 10^{15}$ Hz for the parameter choice here. The hard spectrum of the SSC emission below its SED peak reflects that of synchrotron radiation below $\sim 10^{15}$ Hz ($\propto \nu^{4/3}$). The energy of photons corresponding to the SSC peak is higher than that estimated in Eq. (\ref{eq:icspeak}) due to $\gamma_m > \gamma_c$; it can be estimated by replacing $\gamma_c$ by $\gamma_m$ in Eq. (\ref{eq:icspeak}) in the Thomson regime. However, here, since the KN effect works at the energy of the peak estimated on the assumption of the Thomson regime, the energy of the SSC peak is lower than the prediction and therefore the flux of the SSC emission is suppressed.

\begin{figure}
\includegraphics[clip,width=0.95\linewidth]{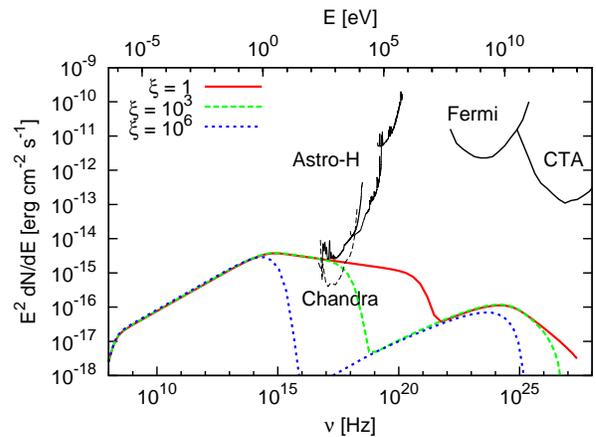}
\caption{Same as figure \ref{fig:eta}, but dependence on the gyrofactor $\xi$.}
\label{fig:xi}
\end{figure}

The roll-off frequency of synchrotron radiation mainly depends on the gyrofactor $\xi$. Thus, the detection of the spectral roll-off constrains the gyrofactor $\xi$. Figure \ref{fig:xi} shows the dependence of the NSBMR SED at $t = t_{\rm dec}$ and $D = 100$ Mpc on the parameter $\xi$. In all the cases electrons are mainly cooled by synchrotron radiation in the slow cooling regime, and therefore the roll-off frequency is governed by synchrotron cooling, i.e., $\nu_{\rm syn}^{\rm max} \propto \xi^{-1} \beta^2$ (see Eq. \ref{eq:syn_max}). There are the cases that NSBMRs do not radiate X-rays because of the low maximum Lorentz factor of electrons (the large gyrofactor). The figure indicates that \emph{Chandra} is sensitive to the parameter $\xi$ in the range of $10^2 \lesssim \xi \lesssim 10^4$. Also, the X-ray telescopes cannot detect NSBMRs in the cases of $\xi \gtrsim 10^4$.

\subsection{Conditions to be detected in gamma-ray bands}

\begin{figure}
\includegraphics[clip,width=0.95\linewidth]{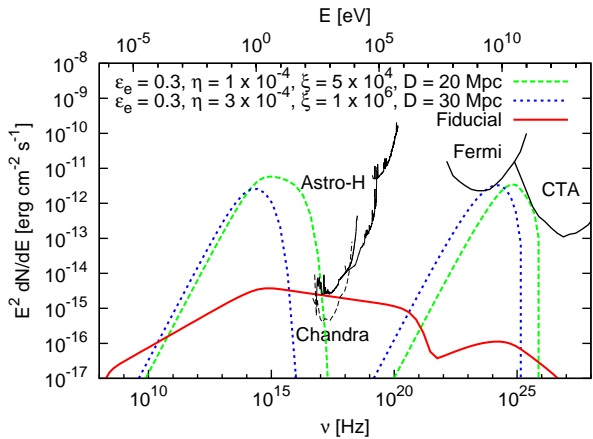}
\caption{Case SEDs where the gamma-ray telescopes can detect NSBMRs at $t = t_{\rm dec}$. The parameters other than the fiducial values are indicated in the legend.}
\label{fig:gamma}
\end{figure}

Let us consider the conditions for gamma-ray telescopes to detect radiation from NSBMRs. The primary cause for the low gamma-ray flux in the fiducial case (figure \ref{fig:fid}) is the low radiative efficiency. A narrow electron spectrum around $\gamma_c$ can enhance gamma-ray flux, which can be realized by a high $\gamma_m$ and a low $\gamma_{\rm max}$. A higher minimum Lorentz factor of electrons at injection $\gamma_m$ increases radiative efficiency due to the absence of radiatively inefficient low-energy electrons. A lower maximum Lorentz factor of electrons $\gamma_{\rm max}$ also increases radiative efficiency at $t = t_{\rm dec}$ by increasing radiation at $t = t_{\rm dec}$. Practically, these correspond to choosing a small $\eta$ and large $\xi$, respectively. These two factors make the bolometric flux close to the maximally achievable bolometric flux $f_{\rm max}$ (Eq. (\ref{eq:fmax})).

A high $\gamma_m$ seems difficult to be realized in the framework of diffusive shock acceleration. Nevertheless, such a high $\gamma_m$ is already required for the SED modeling of some blazars (e.g., \cite{Tavecchio2009MNRAS399L59}). One possible way to realize an electron spectrum with a large $\gamma_m$ is stochastic acceleration downstream of a forward shock where magnetic turbulence is excited \cite{Liu2008ApJ683L163,Fan2010MNRAS406p1337}; the electron spectrum is able to have a high $\gamma_m$ if the escape of particles from the system is inefficient (e.g., \cite{Stawarz2008ApJ681p1725,Lefa2011ApJ740p64}). Detailed discussions on stochastic acceleration in NSBMRs is beyond our scope in this paper. We only point out that there is a motivation to consider narrow electron spectra with high $\gamma_m$.

Another point is the energy of photons where the SED of SSC emission peaks $E_{\rm p,SSC}$, which is mainly related to the detectability by CTA. As estimated in Eq. (\ref{eq:icspeak}), the peak energy of SSC emission is $\sim 40$ GeV under the fiducial parameter set. This is in the \emph{Fermi}-LAT energy range. If $E_{\rm p,SSC}$ increases to the CTA energy range, the flux of both synchrotron radiation and SSC emission decreases. It can be found that $E_{\rm p,SSC}$ depends on all the relevant parameters with negative exponents (see Eq. \ref{eq:icspeak}) and any of these parameters positively correlates with the flux of radiation. Thus, it is impossible to move the SED peak of the SSC emission into the CTA energy range without losing the flux. Furthermore, the KN effect suppresses the SSC flux if the peak frequency is sufficiently high. The condition that the KN affects photons with $E_{\rm p,SSC}$ is $\gamma_c h \nu_{\rm p,syn} \gtrsim m_e c^2$ in the slow cooling regime. Substituting Eqs. (\ref{eq:gc}) and (\ref{eq:syn_brk}) into the condition, one can obtain an inequality $\epsilon_B \lesssim 8 \times 10^{-3} M_{-2}^{-2/5} n_0^{-3/5} \beta_{0.3}^{-4/5}$. A weak magnetic field results in a high $\gamma_c$, and then the KN suppression begins to work at the SED peak of SSC emission. Then, substituting Eq. (\ref{eq:icspeak}) into the inequality, we obtain the condition 
\begin{equation}
E_{\rm p,SSC} \gtrsim 1 \times 10^{11} M_{-2}^{1/15} n_0^{-1/15} \beta_{0.3}^{-1/5} ~~{\rm eV} 
\label{eq:icpeak_limit}
\end{equation}
when electrons are primarily cooled by synchrotron radiation. The very weak dependence on parameters indicates that this condition is robust. Thus, efficient production of gamma rays can happen below $\sim 100$ GeV and the SED peak energy cannot reach $\sim 1$ TeV where CTA is the most sensitive in the differential sensitivity. The projected differential sensitivity of CTA becomes worse monotonically below $1$ TeV and the maximally achievable flux $f_{\rm max}$ is comparable with that at $\sim 200$ GeV. However, the SED peak of SSC emission is difficult to move to $200$ GeV without the reduction of flux. Thus, in order to detect NSBMRs by CTA, a high energy fraction of electrons $\epsilon_e$ and/or their proximity are required under the fiducial parameters for ejecta ($M$ and $\beta$) and the environment of the merger ($n$).

Figure \ref{fig:gamma} shows the cases where CTA and the \emph{Fermi}-LAT can detect NSBMRs at $t = t_{\rm dec}$. Here, as understood from the comparison between the SEDs of synchrotron radiation and SSC emission, the ICS energy-loss of electrons is comparable with or larger than the synchrotron energy-loss. The electrons are in the slow cooling regime.

In the case that the predicted SED reaches the differential sensitivity of CTA, the ICS and synchrotron energy-loss of electrons are comparable, and therefore Eq. (\ref{eq:icspeak}) is applicable, i.e., the energy of photons at the SED peak of the SSC emission is $\sim 40$ GeV. The SED is extended up to $\sim 200$ GeV, and this extension is the target of CTA. CTA can detect gamma rays from NSBMRs within $\sim 20$ Mpc, if such a parameter set is realized. In this case, a very sharp X-ray spectrum, which is detectable, is predicted in soft X-ray ranges because of the maximum energy of electrons high enough to produce gamma rays up to the CTA energy range.

When the radiative efficiency of gamma rays is sufficiently high, the ICS energy-loss of electrons dominates the synchrotron energy-loss, resulting in the SED peak energy of the SSC emission lower than $\sim 40$ GeV. This enables the peak energy to approach $\sim 1$ GeV where the \emph{Fermi}-LAT achieves the best differential sensitivity. In this case, as the gamma-ray radiative efficiency is higher than in the CTA case, the \emph{Fermi}-LAT can detect NSBMRs up to $30$ Mpc. On the other hand, the X-ray telescopes do not always detect them, because it is not necessary for electrons to have energies high enough to emit X-rays. Remember that the \emph{Fermi}-LAT is an all-sky survey in a gamma-ray band. It can search for NSBMRs without {\it a priori} information on their positions. Following Eq. (\ref{eq:nmr}), the expected number of NSBMRs detectable by the \emph{Fermi}-LAT is $0.7 D_{1.5}^3 M_{-2}^{1/3} n_0^{-1/3} \beta_{0.3}^{-1}$ where $D_{1.5} = D / 10^{1.5}$ Mpc, which is close to unity. Thus, NSBMRs might have been already detected as sources without clear identification in other wavelengths.

\begin{figure}
\includegraphics[clip,width=0.95\linewidth]{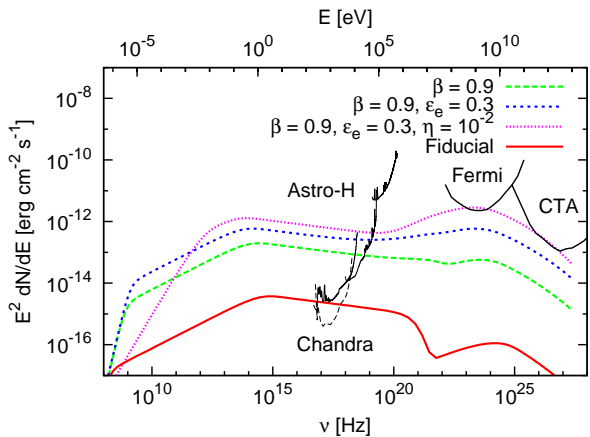}
\caption{SED of NS-NS merger remnants at $t = t_{\rm dec}$ and $D = 100$ Mpc in the cases that the ejecta are powered by magnetically driven winds from rapidly rotating and strongly magnetized massive NSs. The total kinetic energy of ejecta is $7 \times 10^{51}$ erg corresponding to $\beta = 0.9$. The parameters not indicated in the legend are the fiducial values.}
\label{fig:mag}
\end{figure}

So far, we have considered dynamical mass ejection from NSB mergers. However, it is possible that magnetically/neutrino-driven winds in the post-merging phase can additionally powers the ejecta. Here, we examine one of such possibilities; a rapidly rotating and strongly magnetized massive NS produced by a NS-NS merger drives ejecta if the equation of state of NSs is so stiff that the merged NS does not collapse to a black hole (e.g., \cite{Zhang2001ApJ552L35,Metzger2008MNRAS385p1455,Zhang2013ApJ763L22}). The massive NS initially has the rotation energy of 
\begin{equation}
E_{\rm rot} = \frac{1}{2} I \Omega_0^2 = 2 \times 10^{52} I_{45} P_{-3}^{-2} ~~{\rm erg}, 
\label{eq:rotenergy}
\end{equation}
where $I = 10^{45} I_{45}$ g cm$^2$ is the moment of inertia and $P = 10^{-3} P_{-3}$ s is the initial spin period. This huge energy is released within a characteristic time scale, i.e., spin-down time scale, 
\begin{equation}
t_{\rm sd} = \frac{3 I c^3}{B_p^2 R^6 \Omega_0^2} = 2 \times 10^3 I_{45} B_{p,15}^{-2} R_6^{-6} P_{-3}^2 ~~{\rm s}. 
\end{equation}
Here, $B_{\rm p} = 10^{15} B_{\rm p,15}$ G is the strength of the magnetic field at a pole and $R = 10^6 R_6$ cm is the radius of the massive NS. Thus, the total kinetic energy of ejecta can increase by more than one order of magnitude, if the released rotation energy is converted into the kinetic energy before $t_{\rm dec}$ ($> t_{\rm sd}$).

Figure \ref{fig:mag} demonstrates the SED of NS-NS merger remnants in several cases of $\beta = 0.9$ which corresponds to $E = 7 \times 10^{51}$ erg and $t_{\rm dec} = 2$ years. The ejecta expand with mildly relativistic speed of the Lorentz factor of $2.3$. Hence, the calculations with the non-relativistic formalism is marginally justified.

In figure \ref{fig:mag} strong SSC emission hides the roll-off feature of synchrotron radiation in all the cases of $\beta = 0.9$. In the cases of $\epsilon_e = 0.3$ electrons are predominantly cooled by ICS, resulting in the lower cooling break of electron spectra $\gamma_c$ and the lower corresponding SED peak frequency of the synchrotron radiation than $\nu_{\rm p,syn}$. Such a huge energy input into ejecta and the large energy fraction of accelerated electrons enables us to detect gamma rays even for a distance of $D = 100$ Mpc. Thus, the detection of gamma rays from such distant NS-NS merger remnants implies the formation of rapidly rotating and strongly magnetized NSs. The internal pressure of the massive NSs, which depends on the equation of state, should be high enough to support them against gravitational collapse after angular momentum redistribution and neutrino cooling. Thus, the detection of gamma rays may also allow us to constrain the equation of state of NSs with the mass of the massive NSs inferred from GW observations.

The huge energy input allows us to find NS-NS merger remnants by surveys not only in gamma rays by the \emph{Fermi}-LAT but also in X-rays. The Monitor of All-sky X-ray Image (MAXI) is a X-ray telescope on board the International Space Station which monitors all-sky in a soft X-ray band. The sensitivity of MAXI for phenomena with time scale longer than $\sim 1$ year is determined by its confusion limit, $5 \times 10^{-12}$ erg cm$^{-2}$ s$^{-1}$ in the $3$ -- $20$ keV band \cite{Hiroi2009aaxo.conf54,Ueda2009aaxo.conf166}. Thus, although it is not enough to detect NS-NS merger remnants at $D = 100$ Mpc, MAXI can detect and localize them within $\sim 30$ Mpc, in which $\sim 2 D_{1.5}^3 M_{-2}^{1/3} n_0^{-1/3} \beta_{0.9}^{-1}$ NS-NS merger remnants are expected, if the kinetic energy of ejecta is efficiently converted into the energy of electrons ($\epsilon_e = 0.3$). Here, $\beta_{0.9} = \beta / 0.9$. Thus, MAXI might have already detected NS-NS merger remnants as unidentified sources, as well as the \emph{Fermi}-LAT. Future X-ray all-sky monitors such as the Space-based multi-band astronomical Variable Object Monitor (SVOM; \cite{Schanne2010arXiv1005.5008}) may also contribute to the blind searches of NSBMRs in X-ray bands.

The kinetic energy of the mass ejected from a BH-NS merger can also reach $\sim 10^{52}$ erg only by the dynamical mass ejection because the ejected mass can be larger than that in the cases of NS-NS mergers. Note that the observation of GWs can distinguish a NS-NS merger from a BH-NS merger. The dependence of the NSBMR SED on the ejected mass will be discussed in Section \ref{subsec:mandbeta}.

\section{Detectability} \label{sec:detectability}

In the last section we mentioned the detectability of NSBMRs by simply comparing the predicted fluxes with the differential sensitivities of the X-ray and gamma-ray instruments. In reality background emission from galaxies hosting NSBMRs could interfere their identification, unless observational instruments spatially resolve them or they are located outside the host galaxies. Temporal coincidence with GW radiation does not allow background-free observations for NSBMRs because they are long-lasting over years. In this section we discuss the identification of NSBMRs in optical, X-ray, and gamma-ray bands with particular emphasis on possible background emission. The detectability in radio bands is discussed in detail in Ref. \cite{Piran2013MNRAS430p2121}.

Radiation from the active galactic nuclei (AGN) of host galaxies could be a prime competitor to NSBMRs in terms of luminosities. However, they are less important background sources than other sources because AGN bright enough to harbor emission from NSBMRs are rare. Thus, we do not consider AGN as backgrounds.

When the galactic nuclei of host galaxies are not so active like the Milky Way and other low-luminosity AGN, astrophysical objects and/or phenomena in the host galaxies are primary backgrounds for NSBMRs. SNRs, to which a similar model as NSBMRs is applied, are not severe background sources, although the kinetic energy of ejecta is comparable ($E \sim 10^{51}$ erg). The deceleration time $t_{\rm dec}$ of SNRs is $\sim 100$ yr due to the larger mass and slower speed of ejecta. As a result, the flux of a SNR is roughly $10^2$ times smaller than that of a NSBMR. Since a supernova rate is $\sim 10^{-2}$ yr$^{-1}$ for galaxies like the Milky Way, only a few SNRs could contribute to the total flux of SNRs at a time slice. The afterglow of the SGRB associated with a NSBMR is another potential background if SGRBs originate from NSB mergers as have been suggested. Although a SGRB produce highly beamed emission, its afterglow might contribute to the accompanying merger remnant after the significant deceleration of SGRB ejecta even for an off-axis event. However, the total energy of the afterglow is $\sim 10^{48}$ ($\theta_j / 0.1$)$^2$ erg, where $\theta_j$ is the opening angle of the SGRB jet (e.g., \cite{Nakar2007PhR442p166}), which is much smaller than that of the NSB ejecta. Therefore, SGRB afterglows also do not contribute as a background.

Different objects or phenomena dominantly contribute as primary backgrounds in different wavelengths. Although the contents of galaxies depend on their types, the type of galaxies in which NSB mergers predominantly occur is unclear because our knowledge on the mergers is inferred from firmly identified NS-NS binaries in the Milky Way. The observed host galaxies of SGRBs can answer this problem if SGRBs are driven by NSB mergers as believed. However, recent observations have revealed that SGRBs are associated with both early-type (i.e., elliptical) and late-type (i.e., spiral and irregular) galaxies \cite{Berger2009ApJ690p231}, supported by theoretical population studies (e.g., Ref. \cite{Panchenko1999A&AS138p517}). Hence, we consider both possibilities to discuss possible backgrounds. 

To summarize in advance, NSBMRs are expected to be detected in X-ray bands. The identification of the optical emission requires good spacial resolution of telescopes. Gamma-ray emission can exceed possible background emission if its flux is enough to reach the sensitivity of the telescopes.

\subsection{Optical bands}

The superposition of stellar emission is a primary background. Hereafter galaxies without strong active nuclei and prominent starburst activity are called normal galaxies. The typical SED of normal galaxies, including both late-type and early-type galaxies, has a peak at $10^{14}$ - $10^{15}$ Hz originating from the superposition of light of various stars (see \cite{Polletta2007ApJ663p81} for the typical SED of many types of galaxies). In the case of the Milky Way the total luminosity is estimated as $\sim 2 \times 10^{10} L_{\odot}$ in V-band \cite{Gilmore1990book} where $L_{\odot} = 4 \times 10^{33}$ erg s$^{-1}$ is the bolometric solar luminosity. Assuming the typical optical luminosity of a galaxy hosting NSBMRs to be $\sim 10^{10} L_{\odot}$, the flux of the host galaxy in optical bands is $\sim 3 \times 10^{-11} D_{-2}^2$ erg cm$^{-2}$ s$^{-1}$. This flux is even much larger than the optical flux demonstrated in figures \ref{fig:gamma} and \ref{fig:mag} (hereafter, called the high gamma-ray flux cases). A similar discussion is applicable for starburst galaxies.

Good angular resolution of optical telescopes can help to reduce the background by focusing a small part of the galaxy around a NSBMR. Approximating the cross-section of the galaxy to be $\pi R_g^2$, the flux of starlight from a small region around the remnant with the area of $\pi \Delta \phi^2$ is $\sim 8 \times 10^{-14} \Delta \phi_0^2 R_{g,1}^{-2}$ erg cm$^{-2}$ s$^{-1}$ where $\Delta \phi_0 = \Delta \phi / 10^{0}$ arcsec and $R_{g,1} = R_g / 10$ kpc. Therefore, optical telescopes with angular resolution of less than $1^{''}$ such as Keck telescopes \footnote{http://keckobservatory.org}, Subaru telescope \footnote{http://subarutelescope.org}, and HST \footnote{http://www.nasa.gov/hubble/}, can identify synchrotron radiation from the remnant at $D = 100$ Mpc in optical bands in the high gamma-ray flux cases (figures \ref{fig:gamma} and \ref{fig:mag}). In the fiducial case a NSBMR at $D = 100$ Mpc is fainter than the limited magnitude of these telescopes, $\sim 1 \times 10^{-14}$ erg cm$^{-2}$ s$^{-1}$. NSBMRs with $D \lesssim 50$ Mpc are detectable with the better angular resolution of $\lesssim 0.4^{''}$ which is achievable by the HST. A recent observational result that a significant fraction of SGRBs happens in the regions where no stellar light is detected by the HST in their host galaxies \cite{Fong2013arXiv1307.0819} may relax the requirement of angular resolution, if SGRBs originate from NSB mergers. 

\subsection{X-ray bands}

The primary background is X-ray binaries. Low-mass X-ray binaries (LMXBs; $\lesssim 1.5 M_{\odot}$) are associated with old stellar population and are an good indicator of the total stellar mass; LMXBs are known to be the main X-ray sources of early-type galaxies. On the other hand, high-mass X-ray binaries (HMXBs: $\gtrsim 10 M_{\odot}$) are found in star-forming galaxies, being correlating with young stellar population, indicating a star formation rate. Normal late-type galaxies have complex population reflecting complex X-ray source population. If X-ray emission from these binaries are not resolved spatially, the total emission from the binaries becomes a background for the identification of NSBMRs.

X-ray observations of nearby galaxies by \emph{Chandra}, including both early-type and late-type galaxies, and starburst galaxies, resolve X-ray point sources in the galaxies and reveal an empirical relation among the total point-source X-ray luminosity $L_{\rm XP}$, the total stellar mass $M_s$, and star formation rate $\rho$, $L_{\rm XP} = \alpha M_{s,0} + \beta \rho_0$ erg s$^{-1}$ where $M_{s,0} = M_s / 10^0 M_{\odot}$, $\rho_0 = \rho / 10^0 M_{\odot}$ yr$^{-1}$, $\alpha = \left( 1.3 \pm 0.2 \right) \times 10^{29}$ erg s$^{-1}$, and $\beta = \left( 7 \pm 2 \right) \times 10^{38}$ erg s$^{-1}$ \cite{Colbert2004ApJ602p231}. In the case of the Milky Way the total stellar mass of $\left( 6.43 \pm 0.63 \right) \times 10^{10} M_{\odot}$ \cite{McMillan2011MNRAS414p2446} and the star formation rate of $0.68$ - $1.45 M_{\odot}$ yr$^{-1}$ \cite{Robitaille2010ApJ710L11} leads to $L_{\rm XP} \sim 9 \times 10^{39}$ erg s$^{-1}$. Similarly to the Milky Way most of galaxies studied in Ref. \cite{Colbert2004ApJ602p231} have $L_{\rm XP} \sim 10^{39}$ - $10^{40}$ erg s$^{-1}$. Thus, this range of $L_{\rm XP}$ is a reasonable choice to discuss the X-ray background of galaxies without AGN. The X-ray flux from X-ray binaries is estimated as $\sim 8 \times 10^{-15} L_{\rm XP,40} D_2^2$ erg s$^{-1}$ where $L_{\rm XP,40} = L_{\rm XP} / 10^{40}$ erg s$^{-1}$. Hence, NSBMRs at $D = 100$ Mpc can be detected in X-ray bands in the high gamma-ray flux cases as long as the maximum Lorentz factor of electrons $\gamma_{\rm max}$ is high enough to produce X-rays. In the fiducial case they are firmly detectable if the X-ray luminosity of their host galaxies is $L_{\rm XP} \sim 10^{39}$ erg s$^{-1}$. If $L_{\rm XP}$ is close to $10^{40}$ erg s$^{-1}$, their X-ray flux is comparable with the flux of the total flux of X-ray binaries. Even in this case the observations of the temporal evolution of NSBMRs would allow us to identify NSBMRs.

\subsection{Gamma-ray bands}

Galactic diffuse gamma rays are the main background for identification. The diffuse gamma rays primarily originate from cosmic-ray protons accelerated by SNRs. They are produced by cosmic rays propagating in Galactic space via the decay of neutral pions produced by interactions with the ISM. Since star-forming activity is already over in early-type galaxies, diffuse gamma-ray emission from early-type galaxies is not considered.

The diffuse gamma-ray emission of the Milky Way has been estimated by solving the propagation of cosmic rays (e.g., using the GALPROP code \cite{Moskalenko1998ApJ493p694}). Recent calculations required to reproduce the observational result of the \emph{Fermi}-LAT have revealed the diffuse gamma-ray luminosity of the Milky Way is $\sim 7 \times 10^{38}$ erg s$^{-1}$ above 200 MeV in which $\pi^0$ decay of cosmic ray protons dominates \cite{Strong2010ApJ722L58}. Another late-type galaxy M31 is also detected by the \emph{Fermi}-LAT, and its SED can be explained by a scaled SED of the Galactic diffuse emission down to $\sim 4 \times 10^{38}$ erg s$^{-1}$ above 200 MeV \cite{Abdo2010A&A523L2}. Interestingly, Ref. \cite{Abdo2010A&A523L2} found that the number flux of photons above 100 MeV correlates with the star formation rates of Local Group galaxies including several starburst galaxies. Assuming the shape of the SED of these galaxies is similar, this result indicates a correlation between gamma-ray luminosity above 100 MeV and their star formation rates. If this relation is applicable for all normal late-type galaxies and starburst galaxies, the luminosity of diffuse gamma rays is $\sim 10^{40}$ erg s$^{-1}$ even for strong starburst galaxies with star formation rates up to $\sim 20 M_{\odot}$ yr$^{-1}$ such as M82, which indicates the gamma-ray flux of $\sim 8 \times 10^{-15} L_{\gamma,40} D_2^2$ erg cm$^{-2}$ s$^{-1}$ where $L_{\gamma,40} = L_{\gamma} / 10^{40}$ erg s$^{-1}$ is the gamma-ray luminosity. Thus, gamma rays from NSBMRs dominates the diffuse gamma rays of its host galaxy in the high gamma-ray flux cases.

\section{Discussion} \label{sec:discussion}

In Section \ref{sec:results} we discussed the dependence of parameters related to particle acceleration under the fixed parameters on the ejecta of NSB mergers, i.e., their mass $M$ and speed $\beta$, and their environment, i.e., ISM density $n$. However, these parameters are also uncertain at present and should also be determined by observations. The detection of the nine independent characteristic features of NSBMRs allows us to infer all the nine model parameters (see Table \ref{tab:parameters}). In the fiducial case (fig. \ref{fig:fid}) as an example, there are four characteristic frequencies in the SED, i.e., the frequencies of the SED breaks by synchrotron self-absorption and synchrotron cooling ($\nu_{\rm p,syn}$), the roll-off frequency $\nu_{\rm max}$, and the frequency of the SED peak of the SSC emission. Two characteristic fluxes exist, i.e., flux at the SED peaks of the synchrotron radiation and SSC emission. The spectral index of the synchrotron radiation is directly related to the spectral index of electrons $s$. The light curves of NSBMRs, which have not been discussed in detail, can determine $t_{\rm dec}$. Spectroscopic observations of the host galaxies of NSBMRs can determine the distance $D$. The observations of these nine features can in principle determine all the model parameters. Thus, the multi-wavelength observations of NSBMRs are essentially important to understand the physical nature of NSBMRs, NSBs, and their environments. In this section, we provide the dependence of the NSBMR SED on the parameters of the ejecta of NSB mergers ($M$ and $\beta$) and their environment ($n$) in an analytic way to help parameter estimations from observations.

If electrons are accelerated, we expect that cosmic rays are also accelerated in NSBMRs. It is interesting to consider hadronic emission as well as the possible contributions to the observed cosmic rays, regarding NSBMRs as a new class of cosmic-ray accelerators. We also discuss these topics in this section. 

\subsection{SED dependence on ISM density}

The number density of the ISM in the vicinity of NSB merger has been set to be $n = 1$ cm$^{-3}$ throughout this paper. This is motivated by the fact that NS-NS systems firmly confirmed and expected to be merged within a Hubble time, namely B$1913$+$16$ \cite{Hulse1975ApJ195L51}, B$1534$+$12$ \cite{Wolszczan1991Nature350p688}, J$0737$--$3039$ \cite{Burgay2003Nature426p531,Lyne2004Sci303p1153}, and J$1756$--$2251$ \cite{Faulkner2005ApJ618L119}, are located in the disk of the Milky Way in which the typical density of the ISM is $\sim 1 $ cm$^{-3}$. On the other hand, there is a possibility that NS-NS mergers occur in smaller densities because 1) there is a NS-NS binary in a globular cluster, namely B$2127$+$11$C \cite{Anderson1990Nature346p42}, and 2) the ambient ISM density of a significant fraction of SGRBs has been inferred to be $\lesssim 0.01$--$0.1$ cm$^{-3}$ (e.g., \cite{Soderberg2006ApJ650p261,Fong2011ApJ730p26,Fong2012ApJ756p189,Berger2013ApJ765p121}). Also, there is no observational information on the ISM density around BH-NS merger. Thus, it is useful to give the scaling laws of SED on the ISM density $n$.

When electrons are in the slow cooling regime and are predominantly cooled by synchrotron radiation, the frequency of the synchrotron peak $\nu_{\rm p,syn}$ is moved following $\propto n^{-5/6}$ (see Eq. \ref{eq:syn_brk}) and the flux of synchrotron radiation at the peak is scaled proportional to $n^{(2s-3)/3}$. The flux becomes lower in a lower ISM density environment. We can estimate or constrain the ISM density in the vicinity of the NSBs through these dependence. This also allows us to test the NSB hypothesis of the SGRB origin by comparing the inferred density with the circumburst density of SGRBs.

\subsection{SED dependence on ejecta properties} \label{subsec:mandbeta}

According to simulations, the parameters of ejecta, i.e., $M$ and $\beta$, depend on the equation of state of NSs which is uncertain \cite{Hotokezaka2013PRD87p024001}. The mass of ejecta $M$ may be larger than $\sim 10^{-4} M_{\odot}$ up to $\sim 10^{-2} M_{\odot}$ and $\beta$ is distributed from $\sim 0.1$ to $\sim 0.3$ for NS-NS mergers \cite{Hotokezaka2013PRD87p024001}. The ejected mass may be more for BH-NS mergers \cite{Kyutoku2013arXiv1305.6309}. A recent observation of GRB 130603B implies $M \sim 10^{-2} M_{\odot}$ and $\beta \sim 0.1$ if the post-burst radiation is interpreted as a macronova. When the synchrotron cooling of electrons is dominant in the slow cooling regime, the frequency of the synchrotron peak $\nu_{\rm p,syn}$ is scaled according to $\propto M^{-2/3} \beta^{-1}$ (Eq. \ref{eq:syn_brk}) and its flux at $\nu = \nu_{\rm p,syn}$ is proportional to $M^{s/3} \beta^{3s-3}$. Here, we can learn that the flux of NSBMRs is very sensitive to $\beta$.

\subsection{Cosmic-ray acceleration and hadronic gamma-rays}

An analogy with SNRs indicates that a forward shock in NSBMRs can accelerate cosmic rays as well as electrons. The maximum Lorentz factor of the cosmic rays with their atomic number $Z$ and nuclear mass number $A$ is estimated following the same discussion as electrons, 
\begin{equation}
\gamma_{\rm CR,max} = {\rm min} \left( 
\frac{3 Z e \beta R_{\rm dec} B}{20 \xi A m_p c^2},~
\left[ \frac{9 \pi Z e A^2 m_p^2 \beta^2}{10 \xi \sigma_T 
m_e^2 B} \right]^{1/2} 
\right). 
\end{equation}
Here, the nuclear mass is approximately $A m_p$ and $m_p$ is the proton mass. Regarded as a function of $\xi$, the maximum Lorentz factor is numerically represented as 
\begin{eqnarray}
&&\gamma_{\rm CR,max} = \nonumber \\
&&\left\{
\begin{array}{ll}
3 \times 10^{11} \xi^{-1/2} Z^{1/2} A \epsilon_{B,-2}^{-1/4} 
n_0^{-1/4} \beta_{0.3}^{1/2} & (\xi < \xi_{\rm b,CR}) \\
1 \times 10^8 \xi^{-1} Z A^{-1} 
\epsilon_{B,-2}^{1/2} M_{-2}^{1/3} n_0^{1/6} \beta_{0.3}^2 & (\xi \geq \xi_{\rm b,CR}), 
\end{array}
\right.
\end{eqnarray}
where
\begin{equation}
\xi_{\rm b,CR} = 2 \times 10^{-7} Z A^{-4} 
\epsilon_{B,-2}^{3/2} M_{-2}^{2/3} n_0^{5/6} \beta_{0.3}^3.
\end{equation}
Since the value of $\xi$ is more than unity by definition, the dynamical time scale always limits the maximum Lorentz factor of cosmic rays under our choice of parameters. The maximum Lorentz factor corresponds to $1 \times 10^{17}$ eV for protons and $3 \times 10^{18}$ eV for irons, respectively. Therefore, NSBMRs can be sources of ultra-high-energy cosmic rays above the \emph{knee} up to the \emph{ankle} in the cosmic-ray spectrum.

Since cosmic rays are accelerated, gamma-ray emission with the hadronic origin is an interesting possibility as in the cases of SNRs. Here, let us estimate the flux of hadronic gamma-ray emission through the interaction of accelerated protons with the strongly shocked ISM. Since the protons travel a distance of $\sim c t_{\rm dec}$ at $t = t_{\rm dec}$, the optical depth of $pp$ interactions downstream of the forward shock is $\tau_{pp} \sim 4 n \sigma_{pp} c t_{\rm dec} = 9 \times 10^{-7} M_{-2}^{1/3} n_0^{2/3} \beta_{0.3}^{-1}$, where $\sigma_{pp} \approx 5 \times 10^{-26}$ cm$^{-2}$ is the cross section of an inelastic $pp$ collision. The maximally achievable bolometric flux is $\tau_{pp} \xi_{\gamma} \epsilon_p E / 4 \pi D^2 t_{\rm dec} \sim 4 \times 10^{-20} \epsilon_{p,-1} \xi_{\gamma,-1} \epsilon_{p,-1} M_{-2} n_0 D_2^{-2}$ erg cm$^{-2}$ s$^{-1}$, where $\epsilon_{p,-1} = \epsilon_p / 10^{-1}$ is the energy fraction of accelerated protons to the total kinetic energy of ejecta and $\xi_{\gamma,-1} = \xi_{\gamma} / 10^{-1}$ is the energy conversion efficiency of protons to gamma rays in a $pp$ collision. Thus, gamma rays from $pp$ collisions do not compete with the leptonic gamma rays at $t = t_{\rm dec}$. The relative contribution of gamma rays from $pp$ collisions increases at later time because $\tau_{pp}$ is proportional to the time of the system as long as protons are confined in the system while high-energy electrons are cooled. Also, we expect that the flux of high-energy neutrinos is similar to that of the hadronic gamma rays.

\section{Summary} \label{sec:summary}

We have shown that electrons accelerated at forward shocks in NSBMRs can produce high energy emission in X-rays and gamma-ray bands. The X-rays are detectable even by the current generation X-ray telescopes at $t = t_{\rm dec}$ under the fiducial parameter set (see Table \ref{tab:parameters}). The detection of the high-energy radiation reveals the acceleration of high-energy particles in NSBMRs. Also, the observation of the radiation allows us to test the NSB merger hypothesis of SGRBs through the comparison between the estimated ISM density surrounding the mergers and the circumburst density inferred by the observations of SGRB afterglows.

We also suggest that NSBMRs are the accelerators of ultra-high-energy cosmic rays beyond the \emph{knee} up to the \emph{ankle} in the cosmic-ray spectrum. We find that the flux of gamma rays originating from these cosmic rays is much lower than that of the leptonic gamma rays at $t = t_{\rm dec}$.

Nearby NSBMRs ($\lesssim 20$ Mpc) may be also detected by the \emph{Fermi}-LAT and CTA if the radiative efficiency of electrons is high enough at $t = t_{\rm dec}$ (see figure \ref{fig:gamma}). In order to enhance the radiative efficiency a narrow electron spectrum centered at the cooling Lorentz factor of electrons $\gamma_c$ is required. Diffusive shock acceleration, believed to work in SNRs, is difficult to predict such a narrow spectrum. Thus, the detection of gamma rays implies an alternative mechanism like stochastic acceleration as motivated by SNRs.

If the ejecta are powered by another source, i.e., a magnetically/neutrino driven wind in addition to the dynamical mass ejection, the flux of radiation from NSBMRs can be enhanced. We demonstrated that NS-NS merger remnants even at $D = 100$ Mpc are detectable in gamma rays, for example, if the ejecta are powered by a rapidly rotating and strongly magnetized massive NS born through the mergers and the radiative efficiency of electrons is high. The detection of such distant NS-NS merger remnants by gamma rays implies the formation of such massive NSs. The massive NSs must be supported by internal pressure. Thus, the detection may also allow us to constrain the equation of state of NSs through the mass of the massive NSs inferred by the observations of GWs. The huge energy input also enables us to detect the merger remnants by X-ray monitors, such as MAXI, as well as the \emph{Fermi}-LAT in the gamma-ray band. The expected numbers of NS-NS merger remnants within the distances where MAXI and the \emph{Fermi}-LAT are sensitive are $\sim 0.7$ and $\sim 2$, respectively. Thus, they might have been already detected NS-NS merger remnants as sources unidentified with typical X-ray/gamma-ray emitters.

We also scrutinized the detectability of NSBMRs against possible competitors. X-rays from NSBMRs can be firmly detected under the condition required for the gamma-ray detection. On the other hand, they could be competitive to those flux of X-ray binaries in the galaxies hosting the merger remnants in the fiducial case if the total luminosity of the X-ray binaries is close to $\sim 10^{40}$ erg s$^{-1}$. In this case the temporal evolution of X-ray flux is useful to identify NSBMRs. The optical emission of the merger remnants is hidden by total stellar emission in the host galaxies. Thus, telescopes with good angular resolution such as the Keck telescopes, the Subaru telescope, and the HST, are required to spatially resolve the merger remnants in the host galaxies to identify them.

\begin{acknowledgments}
We thank N.~Kawai, R.~N.~Manchester, and A.~Mizuta for fruitful discussions. The work of H.T. is supported by JSPS KAKENHI Grant Number 24$\cdot$9375. K.K. is supported by JSPS Postdoctoral Fellowship for Research Abroad. This work is also supported in part by MEXT KAKENHI Grant Number 24103006 and JSPS KAKENHI Grant Numbers 24000004 and 22244030 (K.I.).
\end{acknowledgments}

\end{document}